# TOWARDS MAINTAINABLE PLATFORM SOFTWARE - DELIVERY COST CONTROL IN CONTINUOUS SOFTWARE DEVELOPMENT

Ning Luo[1] and Yue Xiong[2]

[1]Visual Computing Group, Intel Asia-Pacific Research & Development Ltd, Shanghai, China
ning.luo@intel.com

[2]Visual Computing Group, Intel Asia-Pacific Research & Development Ltd, Shanghai, China
andy.xiong@intel.com

## ABSTRACT

*Modern platform software delivery cost increases rapidly as it usually needs to align with many hardware and silicon's TTMs, feature evolvement and involves hundreds of engineers. In this paper, citing one ultra-large-scale software - Intel Media Driver as an example, we analyse the hotspots leading to delivery cost increase in continuous software development, the challenges on our software design and our experiences on software delivery cost shrink against the targeted design enhancements. We expect the identified hotspots can help more researchers to form the corresponding research agendas and the experiences shared can help following practitioners to apply similar enhancements.*

## 1. INTRODUCTION

Modern platform software delivery cost increases rapidly as it usually needs to align with many hardware and silicon's TTM, feature evolvement and involves hundreds of engineers. In this paper, citing one ultra-large-scale software - Intel Media Driver as one example, we analyse the hotspots in continuous software development leading to delivery cost increase, the corresponding software design challenges and our experiences on software delivery cost shrink by targeted design enhancements. We expect the identified hotspots can help more researchers to form the corresponding research agendas and the experiences sharing can help following practitioners to apply similar enhancements.

## 2. DELIVERY COST IN CONTINUOUS DEVELOPMENT - HOTSPOTS AND CHALLENGES

Intel Media Driver is an ultra-large-scale platform software with around 3 million lines of code and supported by over 300 developers. As the bridge between Intel GPU (graphics processing unit) and the ever-changing end to end media usages, Intel media driver is designed for multiple generations' Intel GPU support on top of different OS and API. It is widely used in diverse media usages ranging from client to cloud, against different software stacks. Every year, it has over one hundred software releases for different purposes.

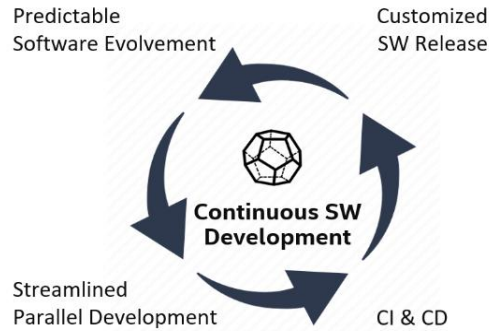

Figure 1.  Delivery Cost in Software Development Cycle

The software delivery cost for Intel Media driver is introduced from various aspects of the continuous software development, as can be shown by Figure 1 above.

**Predictable Software Evolvement**

Execution Predictability is crucial to the long-term success of the large-scale platform software which usually includes the support to dozens of different hardware products.

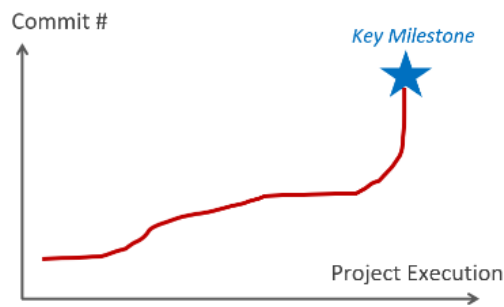

Figure 2.  Predictability Gap in Software Development

As can be shown by Figure 2 above, in the new product enabling, lack of development predictability will lead to extensive rush high risk changes toward cycle's end and frequent product KPIs delay. To those software projects with multiple products enabled/maintained in parallel, the risk will be further amplified because of the unexpected interactions. Finally, we will have to face with more and more unplanned tasks and exponentially increased software delivery cost.

Many factors can lead to the predictability decline in software evolvement. Design perspective, it can be connected to the keep increasing software complexity and gaps on code and effort reusing. Taking Intel Media driver as one example, its software complexity mainly comes from three perspectives -

- **Hardware Complexity**

Intel's GPU portfolio has grown immensely over the last decades. Media driver now needs to support over 50 different media hardware sub-engine IP (Intellectual Property) cores and over 20 different media SOCs (system on chip).

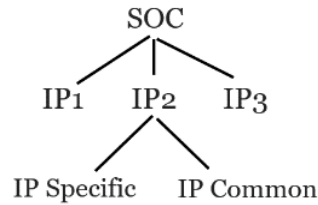

Figure 3. Software Abstraction for Intel GPU

As can be shown by Figure 3. above, Sub-engine IP is the basic hardware unit for Intel GPU. Software abstraction for IP is responsible for GPU workload generation on specific sub-engine. It is the most complex and expensive piece in our new product enabling. Previously, to enable a new version sub-engine IP, we need totally rewrite the corresponding logic, although significant portions of the IP logic is common and can be reused across different IP versions. E.g., the hardware irrelated logic and the hardware features with no changes. It causes not only bigger enabling effort, but also potential risks on the common part divergence between different IPs.

Media SOC is composed of one set of different Media IPs against the SOC specific customizations. For SOCs belonging to the same product family, more than 90% of the logic are common and embodied in the underlying sub-engine IPs. Previously, to enable a new SOC, almost all underlying IPs will be impacted (for the SOC specific programming), while those changes inside underlying IPs can further impact other SOCs belonging to the same product family (based on the same sub-engine IP set). Finally, it will lead to huge code changes impacting hundreds of files together and unexpected big enabling and maintenance effort.

- **OS/API complexity**

Intel Media driver now supports 4 different OSes and 4 different APIs. Although over 85% logic of Media driver is shared between different OS/API, the proportion of the code and effort sharing is much lower. E.g., Changes deemed as OS/ API specific can cause unexpected impacts on other OS/APIs occasionally, so duplicated validation effort seems inevitable.

- **Usage Complexity**

Intel Media driver is widely used in diverse media usages ranging from client to cloud, like Video Playback, Streaming, Gaming, Conference, media delivery and media analytics. For each usage, we need the customized settings on hundreds of different hardware knobs. Without good decoupling, usage specific customizations can easily be mixed up and lead to big debugging and maintenance effort.

In summary, the keep increasing software complexity can ruin the predictability in continuous software development. To mitigate the risk, it requires better code/effort reusing and less mutual impact between software modules in different dimensions - SOCs, IPs, Features, OS/API and Usages.

### Streamlined Parallel Development

Modern platform software often need face with the challenges from parallel development on multiple development paths. Intel Media Driver is one such example. Partial of its source code (mainly for published products' support on Linux) is open sourced and open for contributions from community. While the rest part is kept close source for internal development only. Parallel development is required on both paths. In execution, parallel development can lead to duplicated effort on code change preparation, code review, static analysis, and validation. Even worse, in the long run, code divergence between two code paths turns to be inevitable and the maintenance effort will be multiplied.

To mitigate the risk, it requires a streamlined development model on top of "single code base".

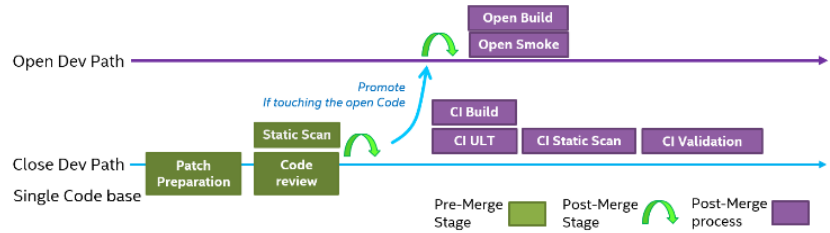

Figure4. Streamlined Parallel Development

The Figure 4 above demonstrates the streamlined development model we used in Intel Media Driver development. The "single code base" will be directly used for internal development while open-source development will be based on its open-source-able subset. After open-source-able subset derived from the single code base, all internal code changes (accounting for around 98% of the total commits) only need to be merged into the "single code base", with one time patch preparation, code review, static code analysis and validation. The corresponding patches to the open-source-able code base will be automatically generated after stripping out the internal only parts. While for the left 2% code commits contributed by community on open-source code base, they will be automatically ported back to the "single code base" for internal review and validation before code merge. In this way, we can best automate and streamline the development flow to minimize the duplicated effort and code divergence on two development paths.

However, to support the streamlined development model, it requires one well organized source code architecture fully aligned with the open-source requirement so that the build system can easily strip out the non-opensource-able part in each code commit with minimal conflict and human intervention.

## Continuous Integration & Continuous Delivery

Continuous Integration (CI) and Continuous Delivery (CD) are the software engineering approaches to promise the reliable software release at any time. They are triggered by each code commit and include series of automatic build and automatic validation stages.

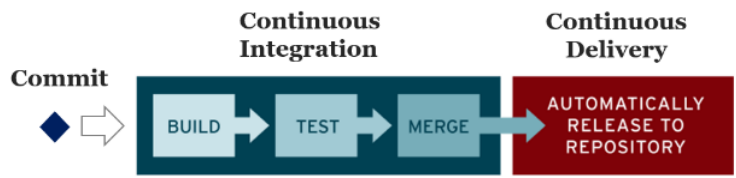

Figure5. Continuous Integration (CI) and Delivery (CD)

To key to a successful CI/CD system is the quick enough developer feedback loop against sufficient build and validation coverage. Still taking Intel Media driver as one example, on average there are around 15 commits per day. It means to promise the best CI efficiency, the CI build and validation stages for each commit need be finished in less than 2 hours. While in real case, because of the large scale of the software, each media driver build needs to include 4 binaries and the corresponding build and Validation stages require around 5 hours. Obviously, build and validation acceleration is critical for the effective CI/CD.

- **Build Acceleration**

Modular build can help to accelerate the build stage. Against modular build, each build request can be broken into smaller build tasks based on sub-modules. In a distributed CI build system, those sub-tasks can be dispatched to multiple machines for parallel build. Meanwhile, some intermediate build results (object files or static libraries) can be reused between several corelated build tasks. Against modular build can help us achieve 3 to 4 times' build time shrink.

But to support modular build, design perspective, our software need be fully decoupled into independent sub-modules.

- **Validation Acceleration**

Smart validation can help the validation stage acceleration which is essentially to choose the just enough validation coverage for each commit based on the code change.

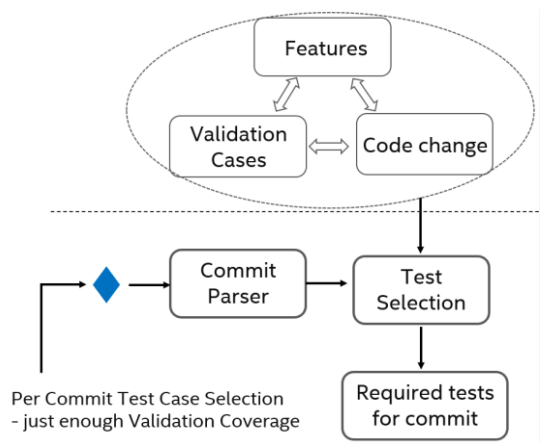

Figure6: Smart Validation Working Flow

Smart Validation infrastructure for Intel Media driver can be shown as Figure6 above. The basis of the smart validation is the traceability from source code to the impacted features and the related test cases. Against the traceability, for each new commit, validation system precisely can easily deduce the most proper validation coverage from the code change. Design perspective, it requires a software design with fully decoupled software features.

**Customized Software Release**

Customized Software Release is essential to large scale software which is the basis for software offering differentiation based on customer needs and IP exposure control on unpublished modules. To support the customized software release. It requires a flexible enough software design for on-demand software tailoring with minimal impact. But to a complex platform software like Intel Media driver, without good decoupling, SOC/IP/feature-based tailoring can easily lead to big changes impacting hundreds of files together. Even worse, the big change can trigger unexpected regressions on irrelated software modules which will lead to bigger validation, debugging and maintenance effort. To fix the gap, design perspective, it requires fully decoupled software modules for different OSes, APIs, hardware platforms, and functional domains.

## 3. OPPORTUNITIES - DESIGN ENHANCEMENT

Effective software delivery Cost Control cannot work without a good software design. We may still take Intel Media driver as one example to analyse how the targeted design enhancements can help on the delivery cost shrink.

Based on the above analysis, we try to have better software decoupling in various dimensions, as can be shown by Figure 7 below.

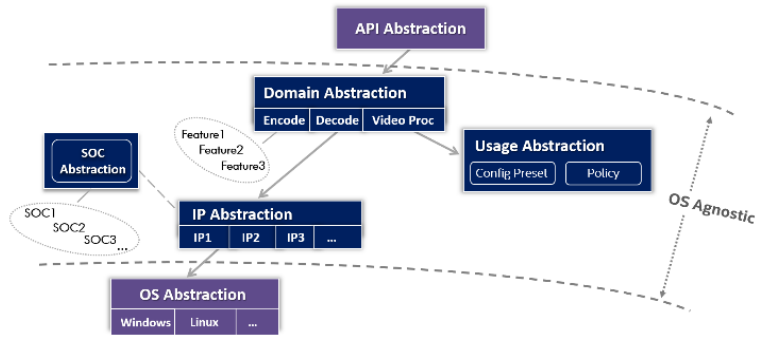

Figure7. Enhanced Software Design

### 3.1 OS and API

In Intel Media driver, only around 15% logic is OS dependent. Two OS dependent modules are introduced to abstract those OS and API differences. At the top, API abstraction module will translate different industry standard APIs into one set of uniform OS agnostic business API. At the bottom, OS abstraction module will abstract services exposed by different OS/KMD into the uniform OS service APIs. With the above two OS dependent modules, the enabling and maintenance effort on the ~85% OS agnostic parts can be better shared across various OS/API. Meanwhile OS/API based source code tailoring can be easily applied against file/folders. It can greatly benefit our parallel development, modular build, and customized software release.

### 3.2 Hardware

- **Abstraction for Sub-engines**

The software abstraction for media sub-engine is decoupled into two layers: One is the hardware irrelated domain abstraction modules mainly responsible for industry standard translation. They can be shared by different sub-engine IPs during our hardware evolvement. The other is hardware related IP abstraction modules responsible for hardware related programming. It will be our focus in the new sub-engine IP enabling. By separating domain abstraction layer from IP abstraction layer, it greatly reduces our enabling and maintenance effort on the around 40% hardware irrelated logic.

- **Abstraction for Features**

The Domain abstraction module and GPU abstraction module are further decoupled into different hardware features and the logic specific to each feature is centralized maintained in the same code block. Feature wise abstraction can help us achieve better traceability from usages/features to source code and greatly benefits the parallel development and smart validation.

- **Abstraction for SOCs**

The SOC specific customizations will be decoupled from the underlying IP and centralized maintained in dedicate SOC abstraction modules. Different SOCs will be fully isolated by files.

In this way, to enable a new SOC, the files impacted can be greatly decreased from over 100 to less than 5 which will greatly reduce our enabling and maintenance effort on dozens of different SOCs.

### 3.3 Usage

Usage based customizations is centralized maintained in usage abstraction modules. Each usage has its own abstraction and different usages will be fully isolated in specific files. In this way, it can avoid interactions between various usages, greatly facilitate the usage specific debugging/tuning and reduce the maintenance effort.

## 4. RESULTS

Combining the aforementioned design enhancements together, it dramatically boosts our engineering efficiency and shrinks the software delivery cost, which can be demonstrated by various engineering measurements –

- The proportion of code commits happened in the last 2 weeks before the important product milestones fell to 8 percent from a 2018 high of 18 percent.
- Average time span for a new commit to be merged into both development paths (open source and close source) fell to 3 days from 4 weeks.
- Average CI developer feedback loop for each code commit was decreased from 5 hours to 1.5 hour.
- Average turnaround time for one new customized software release requirement was reduced from 3 weeks to 2 days.

In total, with the same number of developers, now the team can support 2.5 times more SOCs and IPs comparing with 2018.

## 5. CONCLUSIONS

Delivery cost control is crucial to the success of large-scale platform software. We expect the identified hotspots above can help more researchers to form the corresponding research agendas and the experience shared can provide follow practitioner with insights on similar enhancements.

## ACKNOWLEDGEMENTS

Thanks to all colleagues working on refactoring for continuous software delivery and competency improvement. Appreciate your hard work to turn all our good designs into the reality.

**Authors**

Ning Luo is the senior software architect at Intel. His research interests include software requirements and architecture, continuous delivery, DevOps, and software product lines. Please contact him at ning.luo@intel.com.

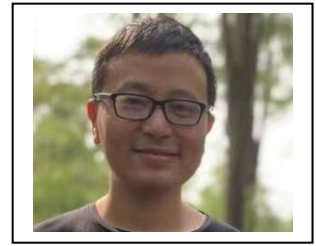

Yue Xiong is the senior director of Intel. Please contact him at andy.xiong@intel.com.

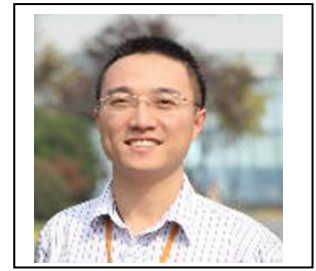